\documentclass[twoside]{article}
\usepackage{Proc_NTSE}
\usepackage{braket}
\usepackage{cite}
\usepackage{psfrag}

\newcommand{\inst}{\hbox{\scriptsize inst}}
\newcommand{\eff}{\text{eff}}
\newcommand{\new}{\hbox{\scriptsize new}}
\newcommand{\ct}{\hbox{\scriptsize ct}}

\begin{document}
\thispagestyle{plain}

\begin{center}

{\Large \bf \strut
 Positronium in Basis Light-Front Quantization
\strut}\\
\vspace{10mm}
{\large \bf 
Paul W. Wiecki, Yang Li, Xingbo Zhao, Pieter Maris
and James P. Vary}
\end{center}

\noindent{% Insert the addresses  here.
\small \it Department of Physics and Astronomy, Iowa State University, Ames, Iowa 50011, USA } \\

%The next command defines running titles:
\markboth{
%Put here the list of authors that will be displayed in running titles:
P. W. Wiecki, Y. Li, X. Zhao, P. Maris and J. P. Vary}
{%Put here the short title of your contribution that will be displayed in running titles:
Positronium in Basis Light-Front Quantization}

\begin{abstract}
We present a calculation of the mass spectrum of positronium within the framework
of the recently developed Basis Light-Front Quantization approach to
non-perturbative quantum field theory. In this calculation, we employ a two-body
effective interaction for the photon exchange, neglecting self-energy effects.
We demonstrate the cancellation of Light-Front small-{\it x} divergences 
within our non-perturbative approach. The resulting spectrum is compared to both
non-relativistic quantum mechanics and previous work in Discretized Light-Cone 
Quantization.
\\[\baselineskip] 
\end{abstract}

\section{Introduction}

The {\it ab initio} calculation of hadron mass spectra and other hadron observables 
remains an outstanding theoretical question. Recent observations of ``tetraquark'' states
in the quarkonium spectrum point to the pressing need for theoretical methods which can address
such systems. In the current ``supercomputing era,'' the computational tools necessary for
such large-scale calculations are now readily available.

The recently developed Basis Light-Front Quantization (BLFQ) \cite{Ref1} approach is a promising tool for 
tackling hadron problems. BLFQ combines the well-known advantages of Light-Front Dynamics \cite{Ref2,hari}
with modern developments in {\it ab initio} nuclear structure calculations, such as the
No-Core Shell Model (NCSM) \cite{Ref3}. The similarity of the Light-Front Hamiltonian formulation to 
non-relativistic quantum mechanics allows the quantum 
field theoretical bound state problem to be formulated
as large, sparse matrix eigenvalue problem. State-of-the-art methods developed for
NCSM calculations can then brought to bear on the hadron problem.

BLFQ has so far been applied to the problem of a free electron in QED. Honkanen \cite{Ref4} and Zhao \cite{Ref5} 
calculated Schwinger's electron anomalous magnetic moment to high precision within the BLFQ approach.
More recently, BLFQ has been applied to time-dependent problems in non-perturbative quantum field theory, such as
non-linear Compton scattering \cite{Ref6}.

Here, we present the first application of BLFQ to a bound state problem, using the positronium system as a test case.
We develop a two-body effective interaction
that acts only on the two-particle sector of the basis. Our calculation is thus equivalent to a ladder truncation
on the Light Front.

\section{Basis Light-Front Quantization}
In BLFQ, hadron observables are calculated by solving the equation
\begin{equation}
P^\mu P_\mu \ket{\Psi}=M^2 \ket{\Psi},
\end{equation}
where $P^\mu$ is the energy-momentum 4-vector operator. Using Light-Cone Gauge, the operator $P^2$ can be constructed.
This operator plays the role of the Hamiltonian in NCSM calculations, and is often referred to as the ``Light-Cone
Hamiltonian'' $H_{LC}\equiv P^2$ \cite{Ref2}.
One can then calculate the matrix elements of this operator in a basis to produce a matrix, which can be diagonalized
to find the mass eigenvalues (squared) and Light-Front amplitudes. The approach is thus non-perturbative.
Since the basis is infinite dimensional, it must be truncated
for the problem to be tractable on a computer. Three separate truncations are made in BLFQ.

The first is truncation on the number of Fock sectors. Since we are solving a quantum field theory, the basis must, in
principle, contain ``sectors'' with all possible numbers and species of particles that can be generated 
by the interactions within $P^2$. The operator $P^2$ contains terms
which change particle number and thus couples the sectors. For example, the ``physical'' positronium state, could
be expressed schematically as
\begin{equation}
\Ket{e^+e^-}_{\hbox{\scriptsize phys}}=a\Ket{e^+e^-}+b\Ket{e^+e^-\gamma}+c\Ket{e^+e^-\gamma\gamma}+d\Ket{\gamma}+f\Ket{e^+e^-e^+e^-}+\cdots.
\end{equation}
In order to have a finite basis, then, we must truncate the Fock sectors at some point. This truncation will be made
by physical considerations. For the moment we restrict ourselves to the $\Ket{e^+e^-}$ and $\Ket{e^+e^-\gamma}$ sectors.
This should be sufficient for generating the Bohr spectrum of positronium. We do not yet make any attempt to 
examine the limit of increasing the number of Fock sectors. 

Secondly, we need a truncation on the Light-Front longitudinal modes. We discretize the longitudinal momentum by putting our
system in a longitudinal box of length $L$ and applying periodic boundary conditions (BCs). Specifically, we choose
periodic BCs for bosons and anti-periodic BCs for fermions. Thus
\begin{equation}
p^+=\frac{2\pi}{L}j,
\end{equation}
where $j$ is an integer for bosons, or a half-integer for fermions. For bosons, we exclude the ``zero modes'', i.e. $j\neq0$. 
In the many-body basis, we select the value of the total longitudinal momentum $P^+=\sum_ip_i^+$,
where the sum is over particles. We then parameterize this using a dimensionless variable $K=\sum_i j_i$ such that 
$P^+=\frac{2\pi}{L}K$. For a given particle $i$, the longitudinal momentum fraction $x$ is defined as
\begin{equation}
x_i=\frac{p_i^+}{P^+}=\frac{j_i}{K}.
\end{equation}

Due to the positivity of longitudinal momenta on the Light Front \cite{Ref9}, fixing $K$ serves as
a Fock space cutoff and makes the number of longitudinal modes finite \cite{1+1}. It is easy to see that $K$
determines our ``resolution'' in the longitudinal direction, and thus our resolution on parton distribution functions.
Real physics corresponds to the limit $K \to \infty$.

Finally, in the Light-Front transverse direction we employ a 2D Harmonic Oscillator (HO) basis. That is, the basis functions
are the eigenfunctions of the potential $V=\frac{1}{2}M\Omega^2\mathbf{r}^2$. Each value of the 
{\it oscillator energy parameter }$b=\sqrt{M\Omega}$ determines a unique complete basis. Convergence rates
depend upon $b$ but the final converged results should not. The basis is made finite
by restricting the number of allowed oscillator quanta according to
\begin{equation}
\sum_i\left(2n_i+|m_i|+1\right)\leq N_{\max},
\end{equation}
where $n$ and $m$ are the radial and orbital quantum numbers of the 2D Harmonic Oscillator, respectively.
Of course, real physics is obtained in the continuum limit of $N_{\max} \to \infty$. Furthermore, we use an M-scheme basis. 
That is, our many-body basis states have a well-defined value of 
\begin{equation}
M_J=\sum_i\left(m_i+s_i\right),
\end{equation}
where $s$ is the helicity. 
These basis states do not, however, have a well-defined value of total $J$.

Since our basis is constructed in single-particle coordinates, the center-of-mass (CM) motion of the system
is contained in our solutions. This problem is also faced in NCSM calculations. The use of the HO
basis combined with the $N_{\max}$ truncation is a great advantage here since it allows for the exact factorization of the wavefunction into ``intrinsic'' 
and ``CM'' components, even within a truncated basis. The CM motion can then be removed from the low-lying spectrum by introducing a 
Lagrange multiplier proportional to $H_{CM}$ (also known as the Lawson term) to the Hamiltonian \cite{caprio}. The extra term
essentially makes CM excitations very costly energetically and thus forces the CM part of the wavefunction to be the ground state
of CM motion. In this way, spurious CM excitations are removed from the spectrum of interest.

It is important to note that, in NCSM calculations, the exact factorization only happens if the isoscalar kinetic energy is used.
That is, the proton and neutron mass are treated as the same. On the Light Front, the kinetic energy can be written as
\begin{equation}
P^+P^-=\sum_{i}\frac{\mathbf{p}_i^2+m^2}{x_i}.
\end{equation}
Comparing to the non-relativistic form $\sum_i\frac{\mathbf{p}_i^2}{2m}$ we see that, on the Light Front, the longitudinal 
momentum fraction $x$ is analogous to mass. Thus the equivalent to using the isoscalar kinetic energy in the NCSM
is for the particles to have {\it equal longitudinal momentum splitting}. For two fermions, 
this situation corresponds to $K=1$ (similarly $K=\frac{3}{2}$ for three fermions).  Indeed, in
initial applications of BLFQ it was found that CM factorization only occured when the total longitudinal momentum
was split equally among the constituents. In order to generalize the 
factorization, the following alternate coordinates were introduced \cite{acta}:
\begin{align}
&\mathbf{q}\equiv\frac{\mathbf{p}}{\sqrt{x}}\nonumber,\\
&\mathbf{s}\equiv\sqrt{x}\mathbf{r}.
\end{align}
When the Hamiltonian is expressed in these coordinates, exact CM factorization is obtained for all eigenstates 
even in a basis with arbitrary distributions of longitudinal momenta as well as an arbitrary numbers of sectors. 
An illustration of the exact CM factorization in BLFQ is given in Refs. \cite{acta,yangli}.

\section{Two-Body Effective Interaction}
We truncate the Fock space to include only $\Ket{e^+e^-}$, and $\Ket{e^+e^-\gamma}$ states. We wish to 
formulate an effective potential acting only in the $\Ket{e^+e^-}$ space that includes the effects 
generated by the $\Ket{e^+e^-\gamma}$ space. In the formalism of effective potentials, we consider the
$P$ space to be the $\Ket{e^+e^-}$ space and $Q$ space to be the $\Ket{e^+e^-\gamma}$ space. Let ${\cal P}$ 
be the operator that projects onto the $P$ space, and ${\cal Q }$ be the operator that projects onto the $Q$ space.

We choose the Bloch form of the effective Hamiltonian. The Bloch form of the effective Hamiltonian has several
advantages compared to the traditional Tamm-Dancoff effective Hamiltonian used in previous studies of positronium
on the Light Front \cite{Ref7}. The Bloch effective Hamiltonian has only unperturbed energies in the energy
denominators, as opposed to an energy eigenvalue which then needs to be found in a self-consistent manner. 
The Bloch Hamiltonian is also automatically 
Hermitian. The Bloch Hamiltonian is given by:

\begin{equation}
\Bra{f}H_{\eff}\Ket{i}=\Bra{f}{\cal P}H{\cal P}\Ket{i}+\frac{1}{2}\sum_{n}\Bra{f}{\cal P}H{\cal Q}\Ket{n}\Bra{n}{\cal Q }H{\cal P}\Ket{i}\left[\frac{1}{\epsilon_i-\epsilon_n}+
\frac{1}{\epsilon_f-\epsilon_n}\right].
\label{eq:heff}
\end{equation}

Here, $H=H_{LC}=P^2$ is the Light-Cone Hamiltonian introduced above.
States $i$ and $f$ are states in $P$ space ($\Ket{e^+e^-}$), while state $n$ is in the $Q$ space ($\Ket{e^+e^-\gamma}$).
$\epsilon_i$ is the unperturbed energy of state $i$, etc. Note that if $i=f$ this reduces to the usual formula from second-order
energy shift in perturbation theory. Furthermore, note that, due to the definition of $H_{LC}$, both the ``Hamiltonian'' and the ``energy'' have
mass-squared dimensions. The mass eigenvalues are thus the square root of the eigenvalues of $H_{LC}$. 
The derivation of \eqref{eq:heff}, based on a perturbative expansion of the Okubo-Lee-Suzuki
effective Hamiltonian \cite{leesuzuki1,leesuzuki2,leesuzuki3,leesuzuki4,leesuzuki5,leesuzuki6},
 is given in Ref. \cite{Ref8}. 

${\cal P}H{\cal P}$ is the part of the Hamiltonian that acts within the two-particle space. It contains two pieces. First, it contains the 
two-particle kinetic energy. Secondly, it contains the Light-Front instantaneous photon exchange interaction. Thus it can be expressed
as 
\begin{equation}
{\cal P}H{\cal P}={\cal P}\left(H_0+H_{\inst}\right){\cal P}.
\label{eq:PHP}
\end{equation}
The instantaneous photon exchange interaction $H_{\inst}$ contains a singularity of the form $\frac{1}{(x_1-x_1')^2}$, where $x_1$ ($x_1'$) 
is the longitudinal momentum fraction of the incoming (outgoing) fermion. This singularity is not physical and must be cancelled.

Since we are interested in primarily the effects of repeated photon exchange, we will only include those combinations of 
terms in ${\cal P}H{\cal Q}$ and ${\cal Q}H{\cal P}$ which generate the photon exchange. We neglect the combinations which result in the photon being
 emitted and absorbed by the same fermion. That is, we do not incorporate the fermion self-energy, 
and therefore no fermion mass renormalization is necessary in this model. In addition,
we work with unit-normalized eigenstates and a fixed value of the coupling constant.

In Light-Front {\it S}-matrix perturbation theory $i=f$. In momentum space, the sum in \eqref{eq:heff} reduces to a sum over 
the polarization states of the photon:
\begin{equation}
\sum_\lambda\epsilon_\mu\left(k,\lambda\right)\epsilon^*_\nu\left(k,\lambda\right)=-g_{\mu\nu}+\frac{k_\mu\eta_\nu +
k_\nu\eta_\mu}{k^\kappa\eta_\kappa},
\label{eq:polsum}
\end{equation}
where $\eta^\mu=\left(\eta^+,\eta^-,\eta^\perp\right)=\left(0,2,\mathbf{0}\right)$ is a unit null vector.
The second term in \eqref{eq:polsum} generates a term identical to the instantaneous photon exchange term
of the Light-Front Hamiltonian, but opposite in sign. That is,  a piece of the second term on the RHS of \eqref{eq:heff}
cancels the instantaneous exchange piece ($H_{\inst}$ in \eqref{eq:PHP}) of the first term on the RHS of \eqref{eq:heff}, leaving the effective 
interaction free of Light-Front small-$x$ divergences \cite{Ref2,Ref9}. 

In our non-perturbative calculation $i\neq f$ and the cancellation of small-$x$ singularities does not occur
in general. This leaves the effective potential with an unphysical singularity, and the resulting interaction is unstable
with increasing $K$. The numerical calculation, as a result, does not converge to a finite number in the continuum limit.

To cure this pathology, we introduce a counterterm of the form
\begin{equation}
\Bra{f}H_{\ct}\Ket{i}=-\sum_{n}\Bra{f}{\cal P}H{\cal Q}\Ket{n}\Bra{n}{\cal Q}H{\cal P}\Ket{i}\left[\frac{\left(a-b\right)^2}
{2ab\left(a+b)\right)}\right],
\label{eq:ct}
\end{equation}
where $a=\epsilon_i-\epsilon_n$ and $b=\epsilon_f-\epsilon_n$. The resulting effective potential is 
\begin{eqnarray}
\Bra{f}H^{\new}_{\eff}\Ket{i}&=&\Bra{f}\left(H_{\eff}+H_{\ct}\right)\Ket{i}
\nonumber \\
&=&\Bra{f}{\cal P}H{\cal P}\Ket{i}+
\sum_{n}\frac{\Bra{f}{\cal P}H{\cal Q}\Ket{n}\Bra{n}{\cal Q}H{\cal P}\Ket{i}}{\frac{1}{2}
\left[\left(\epsilon_i-\epsilon_n\right)+\left(\epsilon_f-\epsilon_n\right)\right]}.
\end{eqnarray}
In this form the cancellation of the instantaneous diagram does occur, and $H^{\new}_{\eff}$ is free of unphysical
Light-Front small-$x$ singularities. We note that our choice of counterterm is equivalent to the prescription used
in previous work in Light-Front effective potentials \cite{Ref7}.

By substituting in the terms from the LFQED Hamiltonian, along with the free-field
momentum-space mode expansions, the effective potential can be easily derived, and the cancellation of the instantaneous diagram
verified. The sum over intermediate states is performed in momentum space, before translating the result back to the 
HO basis.

The result, after cancelling the instantaneous interaction, is
\begin{eqnarray}
\Bra{f}H^{\new}_{\eff}\Ket{i}&=&\Bra{f}PH_0P\Ket{i}+
\alpha\frac{\delta^{x_1'+x_2'}_{x_1+x_2}}{K}\sqrt{x_1x_2x_1'x_2'}\int\frac{d^2q_1}{\left(2\pi\right)^2}
\frac{d^2q_2}{\left(2\pi\right)^2}\frac{d^2q_1'}{\left(2\pi\right)^2}\frac{d^2q_2'}{\left(2\pi\right)^2}
\nonumber\\&&\times \frac{\Psi^{m_1}_{n_1}(q_1)\Psi^{m_2}_{n_2}(q_2)\Psi^{m_1'*}_{n_1'}(q_1')\Psi^{m_2'*}_{n_2'}(q_2')
\bar{u}(1')\gamma^{\mu}u(1)\bar{v}(2)\gamma_{\mu}v(2')}
{\frac{x_1-x_1'}{2}
\left[
\left(\epsilon_i-\epsilon_n\right)+
\left(\epsilon_f-\epsilon_n\right)
\right]}
\nonumber\\&&\times \left(2\pi\right)^2\delta^{(2)}\left(\sqrt{x_1}q_1+\sqrt{x_2}q_2-\sqrt{x_1'}q_1'-\sqrt{x_2'}q_2'\right)
\,,
\end{eqnarray}
where $u$ and $v$ are the 4-component Dirac spinors and
\begin{eqnarray}
\epsilon_i-\epsilon_n &=& \frac{x_1q^2_1+m^2}{x_1}-\frac{x_1'q_1'^2+m^2}{x_1'}
-\frac{(\sqrt{x_1}q_1-\sqrt{x_1'}q_1')^2+\mu^2}{x_1-x_1'}
\nonumber,\\
-\left(\epsilon_f-\epsilon_n\right) &=& \frac{x_2q^2_2+m^2}{x_2}-\frac{x_2'q_2'^2+m^2}{x_2'}
-\frac{(\sqrt{x_2}q_2-\sqrt{x_2'}q_2')^2+\mu^2}{x_2-x_2'}.
\end{eqnarray}
($\mu$ is a fictitious photon mass; see below.)
The integral is evaluated using repeated
2D Talmi-Moshinsky (TM) transformations \cite{Talmi}. With the help of these TM transformations, the 
integral can be reduced down to a single 2D integral, which is evaluated numerically.
The details of the calculation will be presented elsewhere \cite{elsewhere}.

The effective potential $H^{\new}_{\eff}$ has one remaining singularity we have not yet discussed. 
In the event that $x_1=x_1'$ and $\mu=0$, the integrations diverge in the low transverse-momentum limit. 
The singularity thus corresponds to 
the case where the photon has zero momentum. The exact same singularity was found within the context of a 
Bloch Hamiltonian on the Light Front in Ref. \cite{Ref8}. The integral has no singularity if $\mu\neq0$. 
This is why we have introduced $\mu$ as a regulator for this physical infrared divergence. Thus, in addition to
examining the limits $K \to \infty$ and $N_{\max} \to \infty$, we must also consider the limit $\mu \to 0$.

\section{Numerical Results}

In non-relativistic Quantum Mechanics, the hyperfine splitting between the $^1S_0$ and $^3S_1$ states of positronium scales as 
$\alpha^4$, where $\alpha$ is the fine structure constant. At physical coupling, the expected hyperfine splitting
and even the binding energy
are then uncomfortably small relative to the precision of our numerical integrals. Since we would like to use
the hyperfine splitting to test our BLFQ results, we use a large coupling of $\alpha=0.3$ to exaggerate both the
binding energy and the hyperfine splitting. We then compare our results not to experiment, but to the predictions
of non-relativistic Quantum Mechanics at this unphysical value of $\alpha$.
This value of $\alpha$ also allows a direct comparison to the  Discretized Light-Cone
Quantization (DLCQ) results of Ref. \cite{Ref7}.

The numerical results were obtained using the Hopper Cray XE6 at NERSC. ScaLAPACK software \cite{scalapack} was used for the diagonalization.
In this particular implementation of BLFQ, the resulting matrix is quite dense. However, in future applications involving 
multiple Fock sectors, the matrix will be extremely sparse. 

\begin{figure}
\centerline{\includegraphics[width=0.96\textwidth]{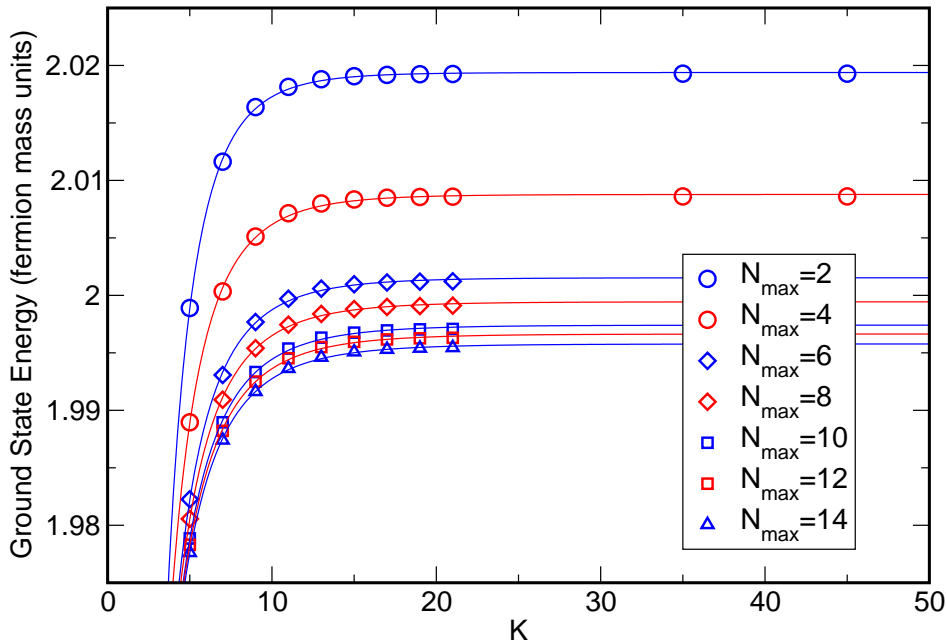}}
\caption{Convergence of the ground state energy with respect to $K$ for various values of $N_{\max}$. The parameters used
are $\alpha=0.3$, $b=0.5m_f$ and $\mu=0.1m_f$. Ground state energy below 2 fermion mass units represents a bound state.}
\label{K}      
\end{figure}
Figure \ref{K} shows the convergence of the
ground state energy as a function of $K$ for various values of $N_{\max}$. In this plot, the basis energy parameter is chosen
to be $b=0.5m$, where $m$ is the fermion mass. We also take $\mu=0.1m$. The same plot made with
a different value of $\mu$ would look qualitatively similar, but with differing absolute energies.
 The ground state energy is also expressed in 
fermion mass units. Thus a ground state energy below $2$ indicates a bound state. The ground state energy is
seen to converge rapidly with increasing $K$. The fitting function used to make the extrapolations is
\begin{equation}
E=a+be^{-c\sqrt{K}}.
\end{equation}
The parameter $a$ is taken to be the result at infinite $K$ for a given $N_{\max}$.

\begin{figure}
\centerline{\includegraphics[width=0.96\textwidth]{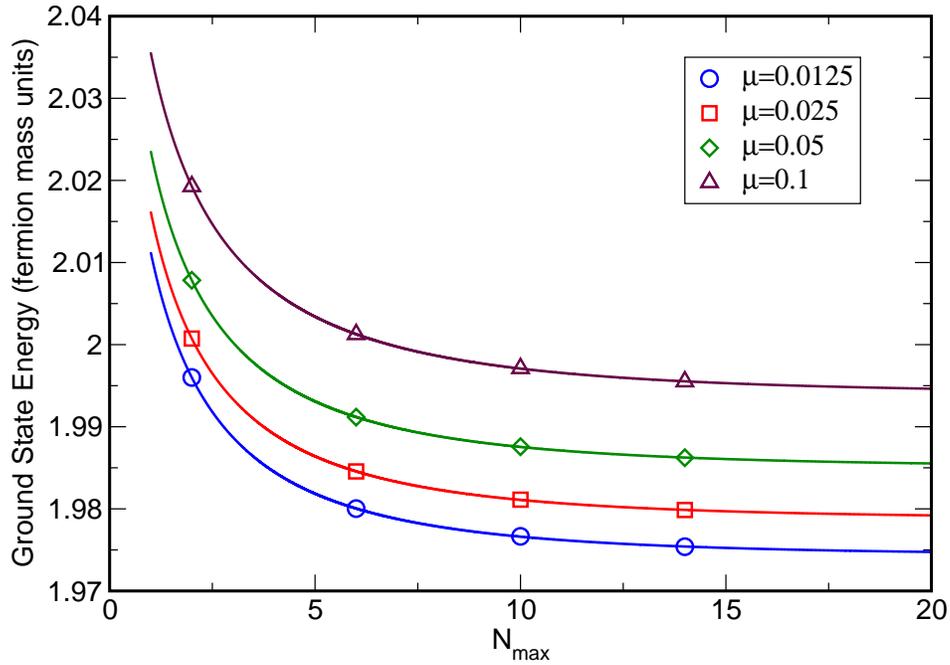}}
\caption{Convergence of the ground state energy with respect to $N_{\max}$ for various values of infrared regulator $\mu$ 
($\alpha=0.3$ and $b=0.5m_f$). Each point has already been extrapolated to the $K \to \infty$ limit as shown in Fig. \ref{K}. }
\label{Nmax}
\end{figure}
We can then plot these extrapolated values as a function of $N_{\max}$. The result is shown in Fig. \ref{Nmax}.
The four curves represent different values of our infrared regulator $\mu$. The ground state energy shows a 
converging trend as a function of $N_{\max}$, although the convergence is slow. In addition, the binding becomes
deeper as we decrease the infrared cutoff $\mu$. For each $\mu$, the curve is fit to the function
\begin{equation}
E=a+be^{-c\sqrt{N_{\max}}}.
\end{equation}
The value of $a$ is then taken to be the ground state energy in the limit $K \to \infty$ and $N_{\max} \to \infty$
for a given value of $\mu$.

\begin{figure}
\psfrag{abc2}{$\sqrt{\mu}$}
\centerline{\includegraphics[width=0.96\textwidth]{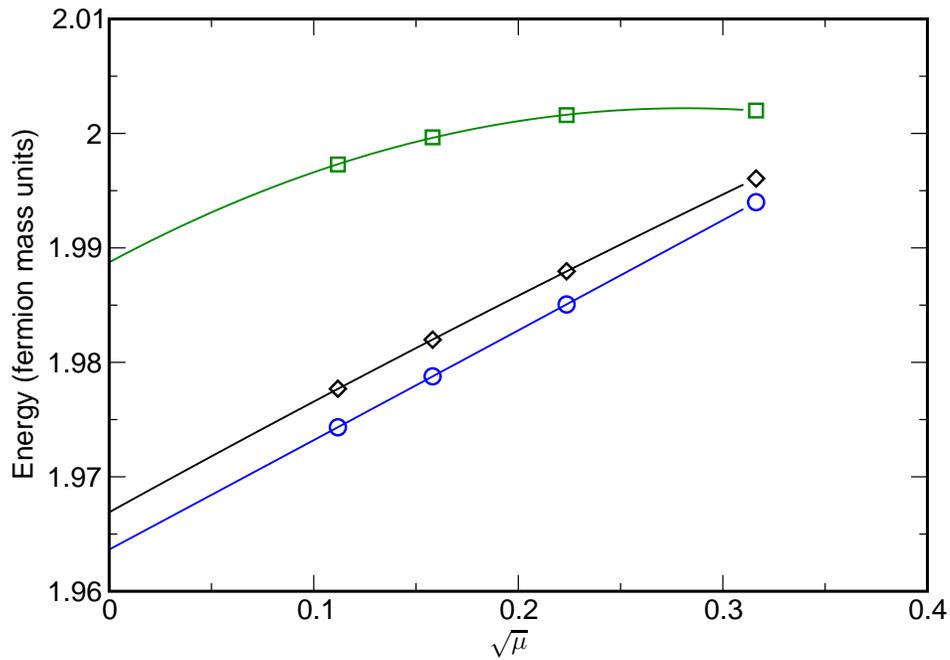}}
\caption{Converged (with respect to $N_{\max}$ and $K$) spectrum as a function of $\sqrt{\mu}$ for $\alpha=0.3$. 
Lowest line is the ground state of $M_J=0$ sector, middle line is the ground state of $M_J=\pm1$ sector and upper line
is the ground state of $M_J=\pm2$ sector }
\label{mu}
\end{figure}
Next we must examine the limit $\mu\to0$. 
The results for the limit $K \to \infty$ and $N_{\max} \to \infty$ are plotted as a function of $\sqrt{\mu}$ in 
Fig. \ref{mu} as the circles. A linear fit is obtained. 
The diamonds represent values for the first excited state, calculated in exactly
the same fashion as the ground state curve. We will compare the splitting between these states to the expected 
hyperfine splitting. 

While we cannot yet calculate the total angular momentum of these states, our identification
of the ground state being a $J=0$ state and the first excited state being a $J=1$ state
is strongly suggested by the following argument. When we do the calculation for $M_J=0$, we see these two states.
If we then do the calculation at $M_J=\pm1$, the lower state disappears and the remaining state is nearly 
(but not identically) degenerate
with the higher state in the $M_J=0$ calculation. Furthermore, both states have disappeared at $M_J=\pm2$. 
This suggests that our ground state has $J=0$ and the first excited state has $J=1$, but we cannot yet prove
this statement.

The squares in Fig. \ref{mu} represent the ground state of our $M_J=\pm2$ calculation. This state disappears
we when we go up to $M_J=\pm3$. This again suggests that this state has $J=2$, but we cannot yet prove it. 
We will then compare this state to the lowest $J=2$ state of the postironium system, which is the $^3P_2$ state.
The curve is fit to a second order polynomial.

\begin{figure}
\centerline{\includegraphics[width=0.98\textwidth]{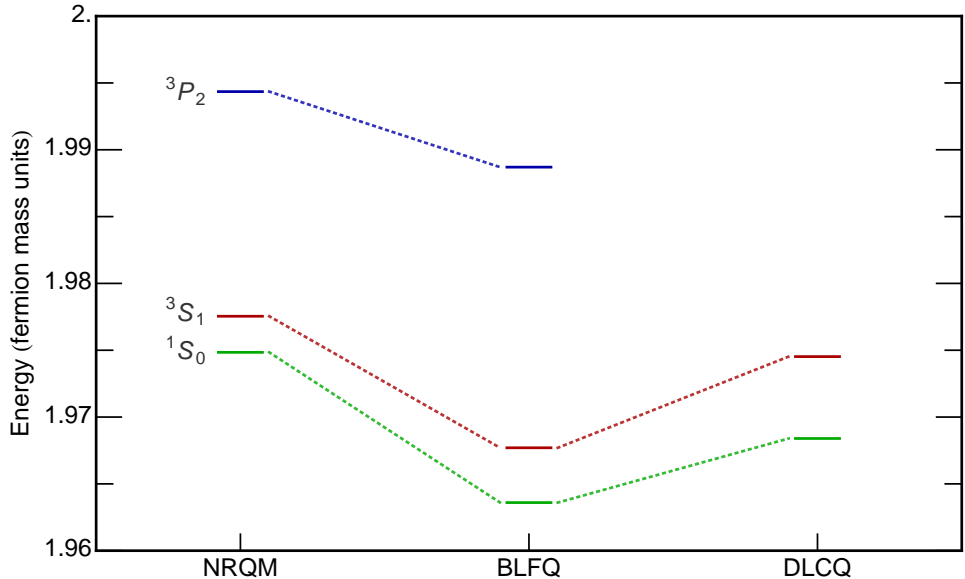}}
\caption{Comparison of BLFQ results to other methods. Quantum Mechanics results are from Ref. \cite{bs}. Results reported
for DLCQ (Ref. \cite{Ref7}) are the result of a Pad\'e extrapolation of a non-converging trend (see Outlook). Dotted lines 
are only to guide the eye; we have not calculated total $J$ for these states in BLFQ (see text).  }
\label{compare}
\end{figure}

The intercepts of the curves in Fig. \ref{mu} with the vertical axis (at $\mu=0$) thus represent the energies in the limit
$K \to \infty$, $N_{\max} \to \infty$ and $\mu\to0$ and can be compared to the predictions of non-relativistic
Quantum Mechanics, and other non-perturbative schemes. This comparison is made in Fig. \ref{compare}. The BLFQ
results are seen to be qualitatively similar to the NRQM expectations, but with a significant overbinding. The hyperfine
splitting is well reproduced. The BLFQ results are also compared to the DLCQ results of Ref. \cite{Ref7}. 
Those authors also find an overall overbinding, and a hyperfine 
splitting of the correct order of magnitude. They do not report a numerical result for the $J=2$ state.  

\section{Summary and Outlook}
We have calculated the spectrum of the positronium system in the non-perturbative Basis Light-Front Quantization
approach.  Instead of tackling the problem directly with a dynamical photon, we have introduced a two-body effective
interaction, which implements the effects of photon exchange, but not the fermion self-energy.  Thus no mass
renormalization was necessary in this calculation. The final converged results agree qualitatively with 
the expectations of NRQM and previous work in DLCQ, with a tendency toward overbinding.

We note that previous authors \cite{Ref7,Karmanov} who have worked on ladder truncation of positronium on the Light Front
have seen a slight dependence on the ultraviolet cutoff of the theory. These authors claim that the divergence they see
will be cancelled when crossed ladder graphs are included in the interaction kernel. While currently we see no evidence of 
such a divergence, we accept that it is present and believe that we are simply not yet at high enough $N_{\max}$ to be sensitive to it.
These issues will be explored in future work.

Our two-body effective potential model should also be applicable to heavy quarkonia if we 
include a confining potential, such as the one motivated by ``soft wall'' AdS/QCD \cite{adsqcd1,adsqcd2,adsqcd3}.
The effective interaction implemented here could then be interpreted as providing a first correction to the basic AdS/QCD spectrum.

Implementation of the problem with one or more dynamical photons in the basis requires the implementation of a
non-perturbative renormalization scheme, such as the Fock Sector dependent scheme of Karmanov {\it et al}
\cite{sectordep}. In addition, the cancellation of unphysical Light-Front singularities would need to occur numerically
within the matrix diagonalization, and not analytically as is done here. The full potential of BLFQ will be 
realized only when these difficulties are overcome. 

%%%%%%%%%%%%%%%%%%%%%%%%%%%%%%%%%%%%%%%%%%%%%%%%%%%%%%%%%%%%%%%%%%%%%%%%%%%%%
\section*{Acknowledgements}

We thank Stanley J. Brodsky, Heli Honkanen and Dipankar Chakrabarti for fruitful discussions.
This work was supported in part by the Department
of Energy under Grant Nos. DE-FG02-87ER40371 and
DESC0008485 (SciDAC-3/NUCLEI) and by the National
Science Foundation under Grant No. PHY-0904782.
A portion of the computational resources
were provided by the National Energy Research Scientic
Computing Center (NERSC), which is supported by the Office of Science 
of the U.S. Department of Energy under Contract No. DE-AC02-05CH11231.

\end{document}